%% file: main.tex
\documentclass{article}
\usepackage[a4paper,left=2cm,top=1.5cm,bottom=1.5cm,right=2cm]{geometry}
\usepackage[utf8]{inputenc}

\usepackage{color}
\usepackage{xcolor}
\usepackage[hyphens]{url}
\usepackage{hyperref}
\usepackage[hyphenbreaks]{breakurl}
\definecolor{linkcolour}{rgb}{0,0.2,0.6}
\hypersetup{colorlinks,breaklinks,urlcolor=linkcolour, linkcolor=linkcolour, citecolor=blue}
\usepackage[onehalfspacing]{setspace}
\usepackage{graphicx}
\usepackage{caption}
\usepackage{subcaption}
\usepackage{multirow}
\usepackage{setspace} 
\usepackage{placeins}
\usepackage{booktabs}
\usepackage{authblk}
\usepackage{lineno}
\usepackage{afterpage}

\makeatletter         
\renewcommand\maketitle{
{
\begin{center}
{\huge \setstretch{1.2}\@title \par}
\vspace{1cm}
{\large \textbf{The ECFA Early Career Researcher's (ECR) Panel}}
\\[1cm]
{\large \@date}
\\[1cm]
\begin{minipage}{0.82\textwidth}
\normalsize The European Committee for Future Accelerators (ECFA) Early Career Researcher's (ECR) panel, which represents the interests of the ECR community to ECFA, officially began its activities in January 2021.  In the first two years, the panel has defined its own internal structure, responded to ECFA requests for feedback, and launched its own initiatives to better understand and support the diverse interests of early career researchers.  This report summarises the panel composition and structure, as well as the different activities the panel has been involved with during the first two years of its existence.
\end{minipage}
\end{center}
\vspace{1.5cm}
\begin{flushleft}
{The ECFA Early Career Researcher's (ECR) Panel: \href{mailto:ecfa-ecr-organisers@cern.ch}{ecfa-ecr-organisers@cern.ch}\\[0.5cm]\@author}
\end{flushleft}}}
\makeatother

\RequirePackage[style=numeric,backend=biber,sorting=none]{biblatex}
\begin{document}

\title{The ECFA Early Career Researcher's Panel: \\composition, structure, and activities, 2021 -- 2022}

\date{\today}

\input{authors.tex}

\maketitle

\clearpage
\section{Executive Summary}
\input{summary.tex}

\section{Panel Structure}
\label{sec:panelstructure}
\input{structure.tex}

\section{Panel Working Groups}
\label{sec:wgs}
\input{wgs.tex}

\section{Summary of Community Interactions}
\label{sec:communityint}
\input{community.tex}

\section{Outlook}
\label{sec:outlook}

\input{future.tex}


\addcontentsline{toc}{section}{Bibliography}
\printbibliography[title=References]

\end{document}

%% file: authors.tex

\author[1]{Andrei~Alexandru~Geanta}
\author[2]{Chiara~Amendola}
\author[3]{Liliana~Apolinario}
\author[*,4]{Jan-Hendrik~Arling}
\author[5]{Adi~Ashkenazi}
\author[6]{Kamil~Augsten}
\author[*,7]{Emanuele~Bagnaschi}
\author[8]{Evelin~Bakos}
\author[5]{Liron~Barak}
\author[3]{Diogo~Bastos}
\author[9]{Giovanni~Benato}
\author[7]{Bugra~Bilin}
\author[10]{Neven~Blaskovic~Kraljevic}
\author[11]{Lydia~Brenner}
\author[12]{Francesco~Brizioli}
\author[13]{Antoine~Camper}
\author[14]{Alessandra~Camplani}
\author[*,15]{Xabier~Cid~Vidal}
\author[16]{Hüseyin~Dag}
\author[11]{Flavia~de~Almeida~Dias}
\author[11]{Jordy~Degens}
\author[17]{Eleonora~Diociaiuti}
\author[7]{Laurent~Dufour}
\author[*,18]{Katherine~Dunne}
\author[19]{Filip~Erhardt}
\author[1]{Stefan-Alexandru~Ghinescu}
\author[7]{Loukas~Gouskos}
\author[20]{Andrej~Herzan}
\author[*,21]{Viktoria~Hinger}
\author[22]{Bojan~Hiti}
\author[*,23]{Armin~Ilg}
\author[24]{Adrián~Irles}
\author[25]{Kateřina~Jarkovská}
\author[8]{Jelena~Jovicevic}
\author[26]{Lucia~Keszeghova}
\author[*,27]{Henning~Kirschenmann}
\author[28]{Suzanne~Klaver}
\author[29]{Sotiroulla~Konstantinou}
\author[30]{Magdalena~Kuich}
\author[31]{Neelam~Kumari}
\author[6]{Katarína~Křížková~Gajdošová}
\author[32]{Aleksandra~Lelek}
\author[33]{Jeanette~Lorenz}
\author[3]{Ana~Luisa~Carvalho}
\author[34]{Péter~Major}
\author[35]{Jakub~Malczewski}
\author[17]{Giada~Mancini}
\author[36]{Laura~Martikainen}
\author[37]{Émilie~Maurice}
\author[38]{Seán~Mee}
\author[39]{Vukasin~Milosevic}
\author[14]{Zuzana~Moravcova}
\author[40]{Laura~Moreno~Valero}
\author[41]{Louis~Moureaux}
\author[42]{Heikki~Mäntysaari}
\author[7]{Nikiforos~Nikiforou}
\author[4]{Younes~Otarid}
\author[43]{Michael~Pitt}
\author[1]{Vlad-Mihai~Placinta}
\author[44]{Géraldine~Räuber}
\author[45]{Giulia~Ripellino}
\author[46]{Bryn~Roberts}
\author[47]{Luka~Šantelj}
\author[*,48]{Steven~Schramm}
\author[49]{Mariana~Shopova}
\author[50]{Kirill~Skovpen}
\author[47]{Aleks~Smolkovič}
\author[51]{Gamze~Sokmen}
\author[*,52]{Paweł~Sznajder}
\author[53]{Lourdes~Urda~Gomez}
\author[54]{Abigail~Victoria~Waldron}
\author[*,55]{Sarah~Williams}
\author[*,56]{Valentina~Zaccolo}
\author[7]{Nima~Zardoshti}
\author[57]{Manuel~Zeyen}

\affil[*]{Editor}
\affil[1]{Horia Hulubei National Institute of Physics and Nuclear Engineering, Bucharest-Magurele; Romania}
\affil[2]{IRFU, CEA, Université Paris-Saclay, Gif-sur-Yvette; France}
\affil[3]{Laboratório de Instrumentação e F\'isica Experimental de Part\'iculas - LIP, Lisboa; Portugal}
\affil[4]{Deutsches Elektronen-Synchrotron DESY, Hamburg; Germany}
\affil[5]{Raymond and Beverly Sackler School of Physics and Astronomy, Tel Aviv University, Tel Aviv; Israel}
\affil[6]{Faculty of Nuclear Sciences and Physical Engineering, Czech Technical University in Prague, Prague; Czech Republic}
\affil[7]{CERN, Geneva; Switzerland}
\affil[8]{Institute of Physics, University of Belgrade, Belgrade; Serbia}
\affil[9]{INFN Laboratori Nazionali del Gran Sasso, L'Aquila; Italy}
\affil[10]{MAX IV Laboratory, Lund University, Lund; Sweden}
\affil[11]{Nikhef National Institute for Subatomic Physics and University of Amsterdam, Amsterdam; Netherlands}
\affil[12]{INFN Sezione di Perugia, Perugia; Italy}
\affil[13]{Department of Physics, University of Oslo, Oslo; Norway}
\affil[14]{Niels Bohr Institute, University of Copenhagen, Copenhagen; Denmark}
\affil[15]{Instituto Galego de F\'isica de Altas Enerx\'ias (IGFAE), Universidade de Santiago de Compostela, Santiago de Compostela; Spain}
\affil[16]{Department of Physics, Bursa Technical University, Bursa; Turkey}
\affil[17]{INFN e Laboratori Nazionali di Frascati, Frascati; Italy}
\affil[18]{Department of Physics, Stockholm University, Stockholm; Sweden}
\affil[19]{Physics department, Faculty of science, University of Zagreb, Zagreb; Croatia}
\affil[20]{Slovak Academy of Sciences, Bratislava; Slovakia}
\affil[21]{Paul Scherrer Institut, Villigen; Switzerland}
\affil[22]{Department of Experimental Particle Physics, Jožef Stefan Institute and Department of Physics, University of Ljubljana, Ljubljana; Slovenia}
\affil[23]{Physik-Institut, University of Zürich, Zürich; Switzerland}
\affil[24]{IFIC, Universitat de Val\`encia and CSIC, Val\`encia; Spain}
\affil[25]{Faculty of Mathematics and Physics, Charles University, Prague; Czech Republic}
\affil[26]{Faculty of Mathematics, Physics and Informatics, Comenius University, Bratislava; Slovakia}
\affil[27]{Department of Physics, University of Helsinki, Helsinki; Finland}
\affil[28]{Nikhef National Institute for Subatomic Physics and VU University Amsterdam, Amsterdam; Netherlands}
\affil[29]{University of Cyprus, Nicosia; Cyprus}
\affil[30]{University of Warsaw, Warsaw; Poland}
\affil[31]{CPPM, Aix-Marseille Université, CNRS/IN2P3, Marseille; France}
\affil[32]{Universiteit Antwerpen, Antwerpen; Belgium}
\affil[33]{Ludwig-Maxilimilians-University Munich \& Fraunhofer Institute for Cognitive Systems IKS, Munich; Germany}
\affil[34]{MTA-ELTE Lendület CMS Particle and Nuclear Physics Group, Eötvös Loránd University, Budapest; Hungary}
\affil[35]{Henryk Niewodniczanski Institute of Nuclear Physics Polish Academy of Sciences, Kraków; Poland}
\affil[36]{Helsinki Institute of Physics, Helsinki; Finland}
\affil[37]{Laboratoire Leprince-Ringuet, CNRS/IN2P3, Ecole Polytechnique, Institut Polytechnique de Paris, Palaiseau; France}
\affil[38]{Institute for Physics, University of Graz, Graz; Austria}
\affil[39]{Institute of High Energy Physics, Beijing; China}
\affil[40]{Institut für Theoretische Physik, Westfälische Wilhelms-Universität Münster, Münster; Germany}
\affil[41]{Universität Hamburg, Hamburg; Germany}
\affil[42]{Department of Physics, University of Jyväskylä and Helsinki Institute of Physics, University of Helsinki, Helsinki; Finland}
\affil[43]{Department of Physics, Ben-Gurion University, Beer-Sheva; Israel}
\affil[44]{Institute of High Energy Physics, Austrian Academy of Sciences, Vienna; Austria}
\affil[45]{Department of Physics, Royal Institute of Technology, Stockholm; Sweden}
\affil[46]{Department of Physics, University of Warwick, Coventry; United Kingdom}
\affil[47]{Jožef Stefan Institute, University of Ljubljana, Ljubljana; Slovenia}
\affil[48]{Département de Physique Nucléaire et Corpusculaire, Université de Genève, Genève; Switzerland}
\affil[49]{Institute for Nuclear Research and Nuclear Energy, Bulgarian Academy of Sciences, Sofia; Bulgaria}
\affil[50]{Ghent University, Ghent; Belgium}
\affil[51]{Middle East Technical University, Physics Department, Ankara; Turkey}
\affil[52]{National Centre for Nuclear Research (NCBJ), Warsaw; Poland}
\affil[53]{Centro de Investigaciones Energ\'eticas Medioambientales y Tecnol\'ogicas (CIEMAT), Madrid; Spain}
\affil[54]{Blackett Laboratory, Imperial College London, London; United Kingdom}
\affil[55]{Cavendish Laboratory, University of Cambridge, Cambridge; United Kingdom}
\affil[56]{University of Trieste and INFN, Trieste; Italy}
\affil[57]{Institute for Particle Physics and Astrophysics, ETH Zürich, Zürich; Switzerland}

%% file: summary.tex

The Early Career Researcher's (ECR) panel of the European Committee for Future Accelerators (ECFA)~\cite{ECFAECRPanel} formed in January 2021, following the recommendations of an initial ECR debate in November 2019~\cite{Bethani:2020ovr}, which aimed to provide ECR input to the 2020 update to the European Strategy for Particle Physics.

The ECR panel includes representatives from each ECFA member entity.  It is mandated to discuss all aspects that contribute in a broad sense to the future of the research field of particle physics, with an emphasis on topics of particular relevance to the ECR community. Following the endorsement of the initial members of the panel in November 2020, the first meetings focused on the establishment of a working structure.  The decided-upon structure involves an organisation committee, responsible for organising and chairing meetings as well as handling outside correspondence, and working groups within the panel, which address or discuss particular topics in more detail. A delegation of five members was selected from within the panel as observers to Plenary ECFA (PECFA) meetings, and one of those members was endorsed as an observer to Restricted ECFA (RECFA) meetings.

This report aims to document the activities and achievements of the ECR panel during its first two years. It is structured as follows: Section~\ref{sec:panelstructure} introduces the initial membership of the panel and describes the procedures for selecting the organisation committee and PECFA/RECFA delegates. Overviews of the activities of the working groups are provided in Section~\ref{sec:wgs}, while Section~\ref{sec:communityint} summarises interactions between the ECR panel and the broader community in a variety of contexts. Finally, Section~\ref{sec:outlook} provides an outlook for the panel.

%% file: structure.tex
\subsection{ECR Panel Membership}



The ECR panel, as of December 2022, consists of 75 members representing 28 distinct entities: 27 ECFA member countries, plus CERN.
Each represented entity is allowed to have up to three members on the panel; countries hosting a major laboratory, as defined by being represented in the Lab Director's Group (LDG), are allowed a fourth representative, so long as at least one of the four representatives is from the laboratory in question.

Panel members are selected by the national RECFA representatives, or the CERN RECFA representative, as appropriate.
The selected members are then proposed to PECFA for endorsement: members endorsed by PECFA in November each year begin their term on January 1 of the following year, while members endorsed by PECFA in July each year begin their term on a back-dated date of July 1 of the same year.
The countries and laboratories represented in the ECFA ECR panel, and their representatives as of December 2022, are listed in Table~\ref{tab:currentpanel}.
An up-to-date list of members is maintained on the ECR panel section of the ECFA website~\cite{ECFAECRPanel}.

\begin{table}[h!]
\scriptsize
\begin{center}
\begin{tabular}{ lll }
  \toprule
  Country/Lab & ECR panel members & Position and speciality\\
  \midrule
\multirow{3}{*}{CERN} & Emanuele Bagnaschi & fellow, theory \\
 & Nima Zardoshti & staff, ALICE \\
 & Laurent Dufour & LD research physics position, LHCb \\
\midrule
\multirow{3}{*}{Austria} & Viktoria Hinger & PhD student becoming postdoc, CMS detector R\&D \\
 & G\'{e}raldine R\"{a}uber & PhD student, HEPHY Vienna, Belle-II \\
 & Se\'{a}n Mee & PhD student, pheno. of strongly interacting dark matter \\
\midrule
\multirow{3}{*}{Belgium} & Aleksandra Lelek & postdoc, phenomenology\\
 & Louis Moureaux & postdoc, CMS \\
 & Kirill Skovpen & postdoc, CMS \\
\midrule
\multirow{1}{*}{Bulgaria} & Mariana Shopova & postdoc, CMS \\
\midrule
\multirow{1}{*}{Croatia} & Filip Erhardt & postdoc, ALICE \\
\midrule
\multirow{2}{*}{Cyprus} & Sotiroulla Konstantinou & PhD student, CMS \\
 & Nikiforos Nikiforou & postdoc, ATLAS\\
\midrule
\multirow{3}{*}{Czech Republic} & Kamil Augsten & assistant professor, ATLAS and COMPASS \\
 & Katar\'{i}na Gajdošová K\v{r}\'{i}\v{z}kov\'{a} &  postdoc, ALICE \\
 & Kate\v{r}ina Jarkovsk\'{a} & PhD student, theory \\
\midrule

\multirow{2}{*}{Denmark} & Alessandra Camplani & assistant professor,  ATLAS \\
 & Zuzana Moravcova & PhD student, ALICE \\
\midrule
\multirow{3}{*}{Finland} & Henning Kirschenmann & senior scientist, CMS \\
 & Heikki M\"{a}ntysaari & senior research fellow, HI theory  \\
 & Laura Martikainen & PhD student, CMS \\
\midrule
\multirow{3}{*}{France} & Chiara Amendola & postdoc, CMS \\
 & Neelam Kumari & PhD student, ATLAS\\
 & \'{E}milie Maurice & assistant professor, LHCb\\
\midrule
\multirow{4}{*}{Germany} & Jeanette Lorenz & senior scientist, data science and quantum computing \\
 & Laura Moreno Valero & PhD student, theory \\
 & Jan-Hendrik Arling & postdoc, ATLAS \\
& Younes Otarid & PhD student, CMS \\
\midrule
\multirow{1}{*}{Greece} & Loukas Gouskos & postdoc, CMS \\
\midrule
\multirow{1}{*}{Hungary} & P\'{e}ter Major & PhD student, CMS \\
\midrule
\multirow{3}{*}{Israel} & Adi Ashkenazi & postdoc, MicroBooNE and DUNE\\
 & Liron Barak & senior lecturer, ATLAS and SENSEI \\
 & Michael Pitt & postdoc, Forward physics at LHC and EIC \\
\midrule
\multirow{5}{*}{Italy} & Francesco Brizioli & postdoc, NA62 \\
 & Valentina Zaccolo & assistant professor, ALICE   \\
 & Giovanni Benato & research fellow, experimental neutrino physics \\
  & Eleonora Diociaiuti & postdoc, Mu2e \\
  & Giada Mancini & postdoc, ATLAS \\
\midrule
\multirow{4}{*}{Netherlands} & Jordy Degens & PhD student, ATLAS \\
 & Flavia de Almeida Dias & assistant professor, ATLAS \\
 & Suzanne Klaver & postdoc, LHCb     \\
&  Lydia Brenner & ATLAS staff, Nikhef \\
\midrule
\multirow{1}{*}{Norway} & Antoine Camper & temporary researcher, AEgIS \\
\midrule
\multirow{3}{*}{Poland} & Magdalena Kuich & assistant professor, NA61/SHINE \\
 & Jakub Malczewski & PhD student, LHCb \\
 & Paweł Sznajder & assistant professor, theory \\
\midrule
\multirow{3}{*}{Portugal} & Liliana Apolinario & assistant researcher, phenomenology\\
 & Diogo Bastos & PhD student, CMS \\
 & Ana Luisa Carvalho & PhD student, ATLAS \\
\midrule
\multirow{3}{*}{Romania} & Andrei Geanta & postdoc, ATLAS \\
 & Stefan-Alexandru Ghinescu & PhD student, NA62 \\
 & Vlad-Mihai Placinta & postdoc/electronics engineer, LHCb \\
\midrule
\multirow{3}{*}{Serbia} & Evelin Bakos & PhD student, ATLAS \\
 & Jelena Jovicevic & research professor, ATLAS \\
 & Vukasin Milosevic & postdoc, CMS \\
\midrule
\multirow{2}{*}{Slovakia} & Andrej Herzan & postdoc, ISOLDE \\
 & Lucia Keszeghova & PhD student, ATLAS \\
\midrule
\multirow{3}{*}{Slovenia} & Bojan Hiti & postdoc, ATLAS \\
 & Aleks Smolkovi\v{c} & PhD student, theory \\
 & Luka \v{S}antelj & postdoc, BELLE II \\
\midrule
\multirow{3}{*}{Spain} & Xabier Cid Vidal & postdoc, LHCb/CODEX-b \\
 & Adri\'{a}n Irles & postdoc, ATLAS, CALICE and ILC \\
 & Lourdes Urda Gomez & PhD student, CMS \\
\midrule
\multirow{3}{*}{Sweden} & Katherine Dunne & PhD student, ATLAS \\
 & Neven Blaskovic Kraljevic & postdoc, accelerators \\
 & Giulia Ripellino & PhD student, ATLAS \\
\midrule
\multirow{3}{*}{Switzerland} & Armin Ilg & postdoc, FCC \\
 & Steven Schramm & assistant professor, ATLAS \\
 & Manuel Zeyen & PhD student, low energy precision physics \\
\midrule
\multirow{3}{*}{Turkey} & Bugra Bilin & senior research fellow, CMS  \\
 & H\"{u}seyin Dag & postdoc, CMS \\
 & Gamze Sokmen & PhD student, CMS \\
\midrule
\multirow{3}{*}{UK} & Bryn Roberts & PhD student, ATLAS \\
 & Abbey Waldron & postdoc, DUNE \\
 & Sarah Williams & postdoc, ATLAS \\
\bottomrule

\end{tabular}
\caption{The full list of ECFA ECR panel members, as of December 2022.}
\label{tab:currentpanel}
\end{center}
\end{table}

\afterpage{\clearpage}

When the original set of panel members was endorsed by PECFA in November 2020~\cite{PECFA-Nov2020}, the number of allocated members followed a slightly different set of rules: DESY (Germany) and Frascati (Italy) were originally considered to be separate entities and thus were also allocated up to three members each, rather than only granting one additional representative for the host country.
It was decided in the November 2021 PECFA meeting~\cite{PECFA-Nov2021} that ECR panel members originally appointed via this system would remain part of the panel for the duration of their mandate, and the rule would be enforced only on incoming panel representatives.
This is the reason that Italy currently has five representatives, as two are from Frascati; the number of representatives for Germany has already been resolved due to a panel member ending their involvement.

The official mandate of the ECR panel members was defined and approved by PECFA~\cite{mandate}.
This mandate defines the interactions between the panel and the parent ECFA group, including the allocation of five PECFA observers, one of which is a RECFA observer; these will be discussed further in Section~\ref{subsec:PECFA}.
The mandate also notes that the panel meets infrequently; rather the overall activities are coordinated by an organisation committee discussed in Section~\ref{subsec:OC}, and day-by-day activities proceed in topical working groups, as detailed in Section~\ref{sec:wgs}.
Beyond defining a structure of the ECR panel, the mandate also defines the eligibility criteria for membership in the ECR panel and stresses that ``\textit{members act as individuals, but should be able to represent the views of early-career researchers in particle physics in the country from which they were nominated.}''

The mandate serves as a solid foundation for the panel activities, but it is intentionally concise in order to leave the panel space to self-organise.
The panel members have therefore agreed upon a few further points in order to facilitate the activities of the ECR panel.
In particular, the panel has decided that quorum for any election, endorsement, or otherwise important choice requires at least 50\% of the panel members to have replied to the poll or other associated means of decision-making.
Additionally, the members have decided to hold at least three meetings of the panel each year: one in January, to set the priorities for the year; one in June, to prepare for the July PECFA meeting; and one in September, to prepare for the November PECFA meeting.
There is the option to have an additional fourth meeting in October in case the September meeting is insufficient; this fourth panel meeting was not deemed necessary in either 2021 or 2022.

While it is understood that not all panel members will be able to join any given meeting, the expectation is that all panel members will make a concerted effort to participate in the three/four panel meetings that take place each year.
The large time gap between panel meetings means that individual commitments often vary from one meeting to the next, which would support increased attendance over the course of the year, but this is not always the case: sometimes external commitments, such as coordination roles, necessitate attending recurring meetings on a long-term basis.
In order to reduce the chance of the same people having to routinely miss panel meetings due to such recurring external constraints, the day of the week on which the previous meeting was held is excluded from consideration when identifying the time of the next meeting. 
This methodology has been applied in every instance since the spring of 2021, but it also has its own limitations: it may exclude the day that would work for the largest number of panel members.
This strategy is therefore noted for future reference as to why meetings have been organised this way to date, without implying that it is an optimal approach.

\subsection{ECR Observers to Plenary and Restricted ECFA}
\label{subsec:PECFA}



According to the ECFA ECR panel mandate \cite{mandate}: ``\emph{From among the ECFA ECR Panel members, a delegation of up to five members is assigned by the panel as observers to Plenary ECFA meetings, and one member is assigned by the panel as observer to Restricted ECFA meetings}''. The following section presents the roles of Plenary and Restricted ECFA committees, responsibilities of observers, and the way how they were selected. 

PECFA is to discuss and decide on all ECFA activities, including evaluating reports delivered by working groups, issuing recommendations to outside organisations, endorsing new country/laboratory representatives, and appointing the ECFA Chair and Secretary. PECFA consists of approximately 100 members (for the full list see Ref. \cite{wwwPecfa}), who typically meet two times per year. Each of these meetings consists of two parts: the first is closed and is used for discussions of topics relevant to ECFA functioning, while the second is open and focuses on informing the high energy physics community about recent activities carried out by ECFA.  

RECFA consists of a single representative from each member country/lab in ECFA (for the full list see Ref. \cite{wwwRecfa}). This community serves as an advisory board to help the ECFA Chair and Secretary in shaping the programme of ECFA activities. RECFA members also serve as the main point of contact with their respective local high energy physics communities and authorities. RECFA meets a few times per year, including for country visits. 

The role and responsibilities of the ECR observers to PECFA and RECFA are as follows. The observers take an active part in PECFA and RECFA meetings, representing the broad point of view of the ECR community. They are also responsible for informing PECFA about activities carried out by the ECR panel. In addition, they hold occasional meetings with the ECFA Chair and Secretary, with a purpose of discussing current developments and stimulating the work of the ECR panel. The observers are also responsible for informing the ECR panel about ECFA activities, in particular those that require the direct involvement of the ECR panel or those that should be considered important by ECRs.

The current list of observers is summarised in Table \ref{tab:rpecfaObservers}. The observers were chosen by a panel-defined selection committee, established \emph{ad hoc} among a set of self-nominated candidates. The selection committee was given guidance by the panel on which criteria to use for the selection: there should be no more than one observer from any given country/laboratory, the observers should be gender-balanced and career-balanced (taking into account both job security and research experience), and the observers should represent a diverse set of research backgrounds.  The selection committee therefore ensured that the observers came from different views and backgrounds and thus could represent the many facets of the ECR high energy physics community.

\begin{table}[!h]
\centering
\begin{tabular}{lll} 
\toprule
Name & Country/Lab & Role  \\ \midrule
Eleonora Diociaiuti     & Italy & PECFA Observer \\
Gianluca Inguglia       & Austria & PECFA Observer \\
Henning Kirschenmann    & Finland & PECFA Observer \\
Lydia Brenner           & Netherlands & PECFA/RECFA Observer \\
Pawe{\l} Sznajder       & Poland & PECFA Observer \\
\bottomrule
\end{tabular}
\caption{The list of observers to the Plenary and Restricted ECFA groups.}
\label{tab:rpecfaObservers}
\end{table}

\subsection{ECR Organisation Committee}
\label{subsec:OC}



The ECFA ECR Panel Organisation Committee (OC) is tasked with coordinating the activities of the ECFA ECR panel.
As of December 2022, there are five OC members, listed in Table~\ref{tab:OCmembers}.

\begin{table}[!h]
\centering
\begin{tabular}{ll} 
\toprule
Name & Country/Lab  \\ \midrule
Jan-Hendrik Arling     & Germany/DESY \\
Sarah Williams         & UK \\
Steven Schramm         & Switzerland \\
Valentina Zaccolo      & Italy \\
Xabier Cid Vidal       & Spain \\
\bottomrule
\end{tabular}
\caption{The list of members of the ECFA ECR Panel Organisation Committee.}
\label{tab:OCmembers}
\end{table}

Following a vote of the panel in April 2021, the OC should preferentially consist of between 5 to 9 members, and it was decided that there is no conflict of interest with, nor requirement for, PECFA/RECFA delegates: they are eligible to be on the OC if they are interested, but there does not have to be a PECFA/RECFA representative on the OC.
The same vote also specified the following responsibilities for the OC:
\begin{itemize}
    \item Organisation of the meetings of the ECR panel -- preparing the agenda, soliciting contributions to the agenda, liaising with speakers, liaising with the PECFA/RECFA representatives to check for updates/news, chairing the meetings, taking minutes of the meetings, and handling other meeting-organisation-related duties;
    \item Handling communication with the outside world -- be listed on the publicly contactable ECFA ECR list (\href{mailto:ecfa-ecr-organisers@cern.ch}{ecfa-ecr-organisers@cern.ch}), handle any emails received from external parties, draft and solicit feedback on official emails to be sent on behalf of the panel when making public statements, and help maintain the public-facing website;
    \item Arrange for public talk allocation -- act as a ``speaker's committee'' type body, which allocates talks in case the panel is contacted asking for someone to present on a given topic;
    \item Coordinate the writing of panel reports -- this includes contributions to the bi-annual ECFA newsletters, as well as annual reports, such as this one.
\end{itemize}
All of the above-mentioned responsibilities have been carried out by the OC members, at least once, since the formation of the OC.

%
%

%% file: wgs.tex

The day-by-day activities of the ECR panel proceed within working groups.
As working groups are intended to represent the current interests and priorities of the ECR panel, which will naturally evolve with time, the list of active working groups is dynamic.
It is expected that some of the existing working groups may cease activities in the year(s) ahead, while other new working groups may be created.

\subsection{Detector R\&D}\label{sec:detector_RnD}

In the spring of 2021, the ECFA Detector R\&D Roadmap effort~\cite{roadmap} organised a series of symposia on different topics.  One of the symposia, on ``Training in Instrumentation'', invited a representative from the ECR panel to present the ECR viewpoint~\cite{wwwTF9Symposium}.  The panel formed an ECR Detector R\&D working group in March 2021, which subsequently solicited a broad range of input from the wider ECR community in preparation for this presentation.  The group organised a town hall meeting to gather immediate feedback and learn which questions were of relevance to the community, and then designed and circulated a survey to students, early career faculty, engineers, and those working in industry with a HEP background. The results of this survey were presented at the symposium and received very positive feedback.  Following the symposium, the analysed results of the survey were published~\cite{training-arxiv} in a document endorsed by the ECR panel; this document was also provided as an input to the ECFA Detector R\&D Roadmap process. 



The composition of the working group is shown in Table~\ref{tab:detector_RnD_WG_composition}.  The following sections describe the output of the ECR Detector R\&D working group.

\begin{table}[htbp]
    \centering
    \begin{tabular}{ll}
    \toprule
        Name & Country/Lab \\ \midrule
        Jan-Hendrik Arling & Germany/DESY  \\ 
        Liron Barak & Israel  \\
        Katherine Dunne &  Sweden \\
        Armin Ilg &  Switzerland \\
        Adrian Irles &  Spain  \\
        Magdalena Kuich &  Poland  \\
        Steven Schramm &   Switzerland \\
        Mariana Shopova &  Bulgaria \\
        Sarah Williams & UK  \\ \bottomrule
    \end{tabular}
    \caption{The list of members of the Detector R\&D working group.}
    \label{tab:detector_RnD_WG_composition}
\end{table}

\subsubsection{Town Hall Event}

The working group organized a town hall discussion~\cite{wwwInstrumentationTownhall} and invited early career researchers in instrumentation, engineers, and technical staff to give feedback on the topics most important for them in terms of their professional development, satisfaction in their current roles, and their view of the future of their career prospects. The ideas and feedback collected in this town hall were used to design the following survey.

\subsubsection{Survey} \label{sec:detector_RnD_survey}

A survey about training in instrumentation was developed by the working group. The survey was circulated among email lists of LHC experiments, national early-career researcher lists, and detector R\&D collaboration lists. The advertisement mail explicitly stressed that the survey was targeted at all ECRs, including those who had not yet been involved in instrumentation work, in order to identify barriers preventing ECR involvement in instrumentation. Engineers were also explicitly encouraged to participate. Overall, a total of 473 responses were recorded.

\subsubsection{Report on Training in Instrumentation}

The analysed results of the survey and the town hall discussion were presented at the ECFA Detector R\&D Roadmap Symposium of Task Force 9 on Training~\cite{wwwTF9Symposium}, and a detailed report was published on arXiv and uploaded on CDS~\cite{training-cds}. The most important concerns raised were the lack of appropriate instrumentation training or access thereto, issues of diversity and inclusion such as ``unconscious bias'', unsatisfactory recognition of instrumentation work and training, and insufficient networking opportunities between different experiments for ECRs in instrumentation. An overview of the report was later presented and discussed with the instrumentation community at the 15\textsuperscript{th} Pisa Meeting on Advanced Detectors~\cite{wwwECRsInInstrumentationPisa}. 

Following up on the lack of networking opportunities in instrumentation, the \textit{Networking in Instrumentation WG} was created.



\subsection{Networking in Instrumentation}

The Early Career Instrumentation Forum (ECIF) is a newly created series of events for ECRs working on or interested in instrumentation. The goal of the ECIF is to foster the interaction of young researchers in instrumentation and to provide access to the experience of senior researchers involved in instrumentation.
%
%
The composition of the WG is shown in Table~\ref{tab:networking_in_instrumentation_WG_composition}.

\begin{table}[htbp]
    \centering
    \begin{tabular}{ll}
    \toprule
        Name & Country/Lab \\\midrule
        Jan-Hendrik Arling & Germany/DESY  \\ 
        Liron Barak & Israel  \\ 
        Eleonora Diociaiuti & Italy/Frascati  \\ 
        Katherine Dunne* & Sweden \\
        Armin Ilg* & Switzerland \\
        Adrian Irles & Spain  \\
        Magdalena Kuich & Poland  \\
        Giada Mancini & Italy/Frascati  \\
        Younes Otarid* & Germany/DESY \\
        Mariana Shopova & Bulgaria \\
        Sarah Williams* & UK  \\ \bottomrule
    \end{tabular}
    \caption{Composition of the Early Career Instrumentation Forum WG. A * is used to indicate members who were hosts of the panel discussion.}
    \label{tab:networking_in_instrumentation_WG_composition}
\end{table}


A first event was planned in 2022 and held on October 26: \textit{The Early-Career Instrumentation Forum: Panel discussion and networking event}~\cite{wwwECIF}. From the survey presented in Section~\ref{sec:detector_RnD}, it was clear that career development is an important topic for ECRs in instrumentation, which is why ``Careers in Instrumentation'' was selected as the topic for the first ECIF event. 

The WG invited senior researchers in instrumentation at different career stages in academia and industry for a panel discussion. After this, the WG collected, polished, and grouped questions to establish the general flow of the discussion. These questions were taken both from the WG members and from registered participants, who were able to submit questions before the event. 

The panel discussion took place on Zoom and was moderated by the hosts. Participants could ask live questions via another moderator. After the panel discussion, the participants had the opportunity to directly discuss with the three panelists in breakout rooms, allowing them to gain more insight on the panelists' career paths.

The first ECIF event had 99 registrations, around 45 of whom actually connected and participated. A recording is available on the event page \cite{wwwECIF}. Future iterations of the ECIF could, for example, target specific instrumentation technologies or other topics relevant for ECRs such as training and networking in instrumentation.

\subsection{Career Prospects and Diversity in Physics Programme Joint Survey}

In the start of 2022, two working groups were formed to tackle questions of career prospects of ECRs and the diversity of the physics programme. Independently, both WGs set out to create a survey to be distributed to the ECR community to better understand their needs and what the ECFA ECR panel could do to promote these issues.
After first initial drafts of the respective survey questions, the WGs decided to merge the two surveys, as their scope was rather similar. 
%
%
The composition of the two working groups who jointly organised the ECR survey is shown in Table~\ref{tab:survey_WG_composition}.

\begin{table}[htbp]
    \centering
    \begin{tabular}{ll}
    \toprule
    Name & Country/Lab \\\midrule
    Kamil Augsten & Czech Republic \\  
    Giovanni Benato &  Italy \\  
    Neven Blaskovic Kraljevic &  Sweden \\ 
    Francesco Brizioli & Italy \\  	
    Eleonora Diociaiuti &  Italy/Frascati \\  
    Viktoria Hinger &  Austria \\ 
    Armin Ilg &  Switzerland \\
    Kateřina~Jarkovská &  Czech Republic \\  
    Katarína~Křížková~Gajdošová &  Czech Republic \\  
    Magdalena Kuich &  Poland \\  
    Aleksandra Lelek &  Belgium \\ 
    Louis Moureaux &  Belgium \\ 
    Giulia Ripellino &  Sweden \\ 
    Steven Schramm &  Switzerland \\  
    Mariana Shopova &  Bulgaria \\ 
    Pawel Sznajder &  Poland \\ 	 
    Abbey Waldron &  UK \\ \bottomrule 
    \end{tabular}
    \caption{Composition of the Career Prospects and Diversity in Physics working groups.}
    \label{tab:survey_WG_composition}
\end{table}

\subsubsection{Structure of Survey}

The survey first gathers information about the personal data of the participants and about their current position, affiliation, and duration of contract. The participants are furthermore asked whether they identify as under-represented. Next, the questionnaire collects information about the field of research and affiliations with research groups or collaborations. The subsequent part contains questions on the diversity of the physics program and whether working in different types of environments affects the careers of ECRs. The penultimate section addresses career perspective and planning as well as work-life balance. The survey finishes with questions on recognition and visibility, and open questions, such as what the ECFA ECR panel could provide in order to support the development of the careers of ECRs.

A reduced set of questions was circulated amongst the ECFA national contacts with the aim to enable a comparison between the early-career and senior researchers opinions on key aspects of the survey.

\subsubsection{Circulation and Preliminary Results}

The ECR survey was distributed among larger experiments, national mailing lists (via the ECFA national contacts and/or ECR panel members), and other ECR mailing lists. A total of 684 responses were collected, while the reduced set of questions for the ECFA national contacts was filled by 26 people. The analysis of the surveys is currently ongoing, and a summary of the results is planned to be released, similar to what was done for the training in instrumentation survey and report (Section~ \ref{sec:detector_RnD_survey}). The results will help steer the future efforts of the ECFA ECR panel. 


\subsection{Electron-Ion Colliders}


The Electron-Ion Collider (EIC) will be the next major investment in, and facility for, particle physics in the USA. The physics programme of this new QCD laboratory includes, but is not limited to, studies of hadronic structures and dynamics between partons, which give rise to phenomena such as gluon saturation. The EIC will be built in Brookhaven National Laboratory (BNL). The construction of the facility is planned to begin in 2024; first data-taking is expected in 2032. More information on the physics case, collider, and detector designs can be found in Ref.~\cite{AbdulKhalek:2021gbh}. 

The ECR EIC working group was created within the ECFA ECR panel in order to raise awareness of the new project and to strengthen the links connecting the nuclear and high energy physics communities, particularly at the level of collaboration between ECRs. Some of the members of this working group have joined the JENAA-organised ``Synergies between the EIC and the LHC'' event~\cite{wwwEICLHCKickOffMeeting} and took an active role in the kick-off meeting related to this interdisciplinary effort~\cite{wwwJENAA}. Other activities of the group include a presentation of the EIC project to the ECFA ECR panel and the establishment of communication channels between this panel and the EIC ECR structures. The current composition of the group is shown in Table~\ref{tab:EIC_WG_composition}.

\begin{table}[htbp]
    \centering
    \begin{tabular}{ll}
    \toprule
    Name & Country/Lab \\\midrule
    Kamil Augsten & Czech Republic \\  
    Bugra Bilin & Turkey \\  
    Heikki Mäntysaari & Finland \\  
    Michael Pitt & Israel \\   
    Gamze Sokmen & Turkey \\  
    Paweł Sznajder & Poland \\  
    Valentina Zaccolo & Italy \\  \bottomrule
    \end{tabular}
    \caption{Composition of the EIC working group.}
    \label{tab:EIC_WG_composition}
\end{table}

%% file: community.tex


During the two years of its existence, the ECFA ECR panel has interacted with the larger high energy physics community in several different contexts.
These interactions are fundamental to the success of the panel, as it is intended to represent the ECR community. The panel must therefore understand the priorities of the ECR community, and convey these priorities to other groups.
The following is a summary of the interactions that have taken place between the panel and other groups, sorted chronologically.
Some of these interactions have been mentioned earlier in the report, but are nonetheless repeated here, such that this section represents a full summary of the panel's involvement with the larger community.

The panel was officially formed in January 2021, but the original members were informed of their selection in late November 2020.
This early start is important, as the panel quickly identified a cause that was important to the ECR community: the budget allocated for frontier research in Horizon Europe.
The panel members formulated a statement, determined how the panel would endorse such statements (a quorum of 50\% participation), and publicly issued the endorsed statement on the ECFA website on December 14, 2020~\cite{RescueHorizonEurope}.
This was all done before the official kick-off of the panel because the members were passionate about the topic and needed to act quickly to represent the ECR community.

The next notable interaction relates to the ECFA Detector R\&D Roadmap process, as discussed in Section~\ref{sec:detector_RnD}.
In this instance, the panel received a request from ECFA, in March 2021, to gather feedback from the ECR community on the topic of training in instrumentation.
The panel quickly reacted to this request by holding a live meeting and distributing a survey, which gathered 473 responses.
The resulting information was conveyed back to ECFA in April 2021, and was written up in a report submitted to arXiv in July 2021~\cite{training-cds}.
A follow-up summary was then presented by a panel member at the 15\textsuperscript{th} Pisa Meeting on Advanced Detectors~\cite{wwwECRsInInstrumentationPisa}.

At the end of 2021, the ECR panel was invited to submit a contribution to the ECFA Newsletter \#8 (Winter 2021).
The panel responded to this opportunity, providing a one-page summary of the first year of activities.
Following this first successful contribution, the panel has been invited and has subsequently submitted contributions to the ECFA Newsletter \#9 (Summer 2022) and the upcoming ECFA Newsletter \#10 (Winter 2022), thereby providing bi-annual updates on the progress and activities of the panel.
All of the newsletters can be found in Ref.~\cite{newsletters}.  

The war in Ukraine has had, and is continuing to have, a profound impact on the high energy physics community and society at large.
The situation is delicate, as there can be strong views on how to react; they can present opposing directions, and with a variety of different motivations.
Moreover, the ECR viewpoint on how to best respond to this situation may or may not be aligned with the decisions taken in other contexts, and those decisions can have a significant impact on individuals within the ECR community.
The panel therefore decided it was important to provide ECR input to the CERN Council meeting in June 2022, where a decision was planned on whether or not to suspend the international collaboration agreements between CERN and Russia, Belarus, and JINR.
This input took the form of a brief statement on different possible scenarios, which summarised the results of a panel-internal survey; this survey was internal both due to time constraints and to ensure that each ECFA member country had equal say in the statement.
Due to the sensitive nature of the topic, the statement was not publicly released, rather it was sent to the PECFA group, the CERN Council Secretariat, and the CERN Council Chair.

The first survey that the panel distributed to the ECR community, on the topic of training in instrumentation, was very successful in obtaining useful feedback on the current situation as seen by ECRs and possible future improvements.
Multiple groups within the panel therefore began to prepare subsequent surveys to gather collective feedback on other topics of importance to the ECR community.
These efforts eventually merged, both to provide a single survey rather than tiring out the community by asking for feedback too often, and to benefit from having more people reviewing the survey in advance to ensure that it asked the right questions and that it did so in an appropriate manner.
The resulting survey, which covered a variety of topics relating to career prospects and diversity in physics programme, was distributed in September 2022.
By the time the survey closed in October, 684 responses were received, providing a wealth of data to help understanding the ECR view on numerous important questions.
The analysis of this data is now ongoing, and will be released publicly as soon as the results are ready for distribution.

One of the challenges identified in the training in instrumentation survey was the lack of inter-experimental networking opportunities for ECRs involved in instrumentation.
The panel has been working to remedy this, both by gathering information on relevant schools/conferences~\cite{instSchools} and by organising dedicated events.
The first panel-led event to support ECR networking in instrumentation took place in October 2022, with three
invited speakers/panelists, and 45 participants~\cite{wwwECIF}.
Future events are being planned to continue to support this part of the ECR community.

Anyone from the high energy physics community is always welcome to contact the ECR panel's organisation committee (\href{mailto:ecfa-ecr-organisers@cern.ch}{ecfa-ecr-organisers@cern.ch}) to suggest future activities or other types of interactions that should be pursued.
Anyone who is interested in following the activities of the ECR panel is welcome to sign up to the announcement mailing list (\href{mailto:ecfa-ecr-announcements@cern.ch}{ecfa-ecr-announcements@cern.ch}); this is a CERN-hosted mailing list, and thus registration requires either having a CERN account or CERN lightweight account, the latter of which does not require any CERN affiliation.

%% file: future.tex
In the November 2022 meeting of PECFA, new members of the ECFA ECR panel were appointed to replace members ending their mandate, resulting in a change-over of around half of the ECR panel members. When discussing the logistics of the changeover of the panel (which proceeds through the RECFA delegates of each country/institute), it was noted that the two-year mandate with an option to extend for a further two was potentially less applicable to ECRs (who are often on short-term contracts). There was also a discussion of whether more turnover in the panel (i.e. members only doing two-year terms) would increase opportunities for more ECRs to participate in the panel and benefit both from the professional experience and visibility. However, as significant time was spent in the first year of the panel's activities establishing a working structure, it was agreed that any ECR member wishing to remain for at least another year would have their mandate renewed, with the view that the new panel would revisit this discussion. This approach also ensures some continuity in the organisation committee and existing working groups.

In the September 2022 meeting of the ECR panel, it was decided that a handover meeting would be organised by the outgoing organisation committee to facilitate a smooth transition to the new panel. The incoming panel can benefit from the structures already in place whilst also taking ownership to review and adapt the activities/operation of the panel into the future. This handover meeting will take place in January 2023, and all outgoing and incoming panel members will be invited. In addition to revisiting the strategy for extending the mandate of the panel, the new panel will be encouraged to discuss the ideas/suggestions raised in the September meeting. In particular, there was a discussion on how to further use the ECR panel to facilitate discussions in different countries on the ECFA roadmap following the European Strategy Update. In 2022 two dedicated ECR meetings on future colliders were organised in the UK, but it was noted that expecting the ECR panel members from each country to organise such meetings would be resource intensive and that organising a central ``ECR Future Collider'' event through the panel, with the potential for break-out sessions for specific countries/institutes, may be more efficient. 

The activities of the ECR panel in its first two years of operation have enabled the views, concerns, and ideas of ECRs to influence discussions and decision making across a broad range of areas of high energy physics. To aid the first significant changeover in panel composition, this report has highlighted the main achievements of the panel to date, and has provided a brief look to the future challenges and opportunities for the incoming ECR panel.
